\documentclass[aps,pre,twocolumn, titlepage]{revtex4}
\usepackage{graphicx}
\usepackage{color}
\usepackage{dcolumn}
\usepackage{amsmath}
\usepackage{amsfonts}
\usepackage{bm}
\usepackage{epstopdf}
\begin{document}
\title{Directional locking in a 2D Yukawa solid modulated by a 2D periodic substrate}
\author{Wenqi Zhu$^1$, C. Reichhardt$^2$, C. J. O. Reichhardt$^2$, and Yan Feng$^1$ $^\ast$}
\affiliation{
$^1$  Institute of Plasma Physics and Technology, School of Physical Science and Technology, Soochow University, Suzhou 215006, China\\
$^2$ Theoretical Division, Los Alamos National Laboratory, Los Alamos, New Mexico 87545, USA\\
$\ast$ E-mail: fengyan@suda.edu.cn}
\date{\today}

\begin{abstract}

Directional depinning dynamics of a 2D dusty plasma solid modulated by a 2D square periodic substrate are investigated using Langevin dynamical simulations. We observe prominent directional locking effects
when the direction of the external driving force is varied relative to the underlying square substrate. These locking steps appear when the direction of the driving force is close to the symmetry direction of the substrate, corresponding to the different dynamical flow patterns and the structures. In the conditions between the adjacent locking steps, moving ordered states are observed. Although the discontinuous transitions often occur between the locking steps and the non-locking portion, the continuous transitions are also found around the locking step associated with the disordered plastic flow close to its termini. Our results show that directional locking also occurs for underdamped systems, which could be tested experimentally in dusty plasmas modulated by 2D substrates.

\end{abstract}

\maketitle

\section{Introduction}

Collective dynamics of interacting particles have been widely investigated in various two-dimensional (2D) physical systems, such as colloids~\cite{Reichhardt:2005}, vortex lattice in superconductors~\cite{Harada:1996}, pattern-forming systems~\cite{Reichhardt:2003,Sengupta:2010}, Wigner crystals~\cite{Williams:1991}, and dusty plasmas~\cite{Thomas:1996,I:1996,Chu:1994,Morfill:2009,Thomas:1994,Fortov:2005,Piel:2010,Bonitz:2010,Melzer:1996,Merlino:2004,Feng:2008,Thomas:2004}. The dynamical behaviors
of these systems when modulated by varies kinds of external substrates are
of great interest, as studied in~\cite{Reichhardt:2005,Reichhardt:1999,Reichhardt:2003,Williams:1991}.
These systems can be driven by different types of external forces,
leading to various distinct  dynamical responses~\cite{Reichhardt:2017}.
The driving forces may be either DC~\cite{Reichhardt:2017} or AC~\cite{Reichhardt:2002,Tekic:2010},
with a fixed direction~\cite{Reichhardt:2017} or
varying directions~\cite{Reichhardt:1999,Reichhardt:2012}.
The dynamical responses that have been observed include the pinning and depinning~\cite{Reichhardt:2017, Schwarz:2001}, structural phase transitions~\cite{Mandelli:2015}, and the phase locking effects~\cite{Reichhardt:1999,Reichhardt:2012,Tekic:2010}.

For particles moving over 2D periodic substrates, different dynamical phases can arise due to the symmetry
of the substrates~\cite{Schmiedeberg:2008,Neuhaus:2013,Reichhardt:20123,Xiao:2010, Tierno:2007,Lacasta:2005,Reichhardt:2004}. One interesting  phenomenon related to these dynamical phases is the so-called directional locking~\cite{Xiao:2010, Tierno:2007,Lacasta:2005,Reichhardt:2004}, where the particles become dynamically locked to move in certain directions relative to the substrate symmetry, even when the driving force is not aligned in these directions. That is to say, for the particles driven with varying directions, their motion becomes locked to specific directions relative to the substrate symmetry, instead of following the driving force direction~\cite{Reichhardt:1999,Reichhardt:2012}. Directional locking effects have been widely studied in experiments~\cite{Xiao:2010, Tierno:2007,Marconi:2000,Silhanek:2003,Villegas:2003} and simulations~\cite{Lacasta:2005,Reichhardt:2004,Gopinathan:2004,Pelton:2004,Long:2008,Speer:2010,Balvin:2009,Herrmann:2009,Khoury:2008,Reichhardt:20032,Koplik:2010, Carneiro:2002} in various systems, such as colloids~\cite{Korda:2002,MacDonald:2003,Lacasta:2005,Stoop:2020}, superconducting vortices~\cite{Reichhardt:1999,Reichhardt:2012}, active matters~\cite{Reichhardt:2020}, and microfluids~\cite{Risbud:2014}.
The previous directional locking studies focus on the structure transition of the particle arrangement~\cite{Reichhardt:2012}, while the depth of the substrate and the commensurability ratio vary, for different substrates~\cite{Reichhardt:20122}, such as the triangular periodic substrates~\cite{Reichhardt:2012}, the square periodic substrates~\cite{Reichhardt:1999,Reichhardt:2012}, and the quasicrystalline substrates~\cite{Reichhardt:20122}. In~\cite{MacDonald:2003,Lacasta:2005,Long:2008,Speer:2010}, the directional locking effect is also exploited to sort different kinds of particles. In short, the directional locking effect has already been widely studied in the overdamped and strongdamped systems~\cite{Reichhardt:1999,Reichhardt:2012,Pelton:2004}, so that it would also be interesting to investigate whether this effect also occur in underdamped systems, such as dusty plasmas as studied here.

Laboratory dusty plasma typically refers to the mixture of micron-sized dust particles, free electrons, free ions, and neutral gas atoms~\cite{Thomas:1996,I:1996,Chu:1994,Morfill:2009,Thomas:1994,Fortov:2005,Piel:2010,Bonitz:2010,Melzer:1996,Merlino:2004,Feng:2008,Thomas:2004}. In the typical laboratory conditions, by absorbing free electrons and ions in the plasma, these micron-sized dust particles can be charged to a high negative charge of $\sim -{10}^{-4} e$, interacting with each other through the Yukawa repulsion~\cite{Konopka:2000}. As a result, these dusty particles are strongly coupled, which can be self-organized into a single layer suspension, called 2D dusty plasma~\cite{Feng:2011,Qiao:2014}, exhibiting the typical solid-~\cite{Feng:2008,Hartmann:2014} or liquid-like properties~\cite{Thomas:2004,Feng:2010}. While moving in the plasma environment, dust particles experience a weak frictional gas drag force~\cite{Liu:2003}, with the typical damping rate of $\nu \sim 1{\rm s}^{-1}$~\cite{Feng:2008}. In experiments, dust particles can be directly imaged and then tracked~\cite{Feng:20163}, so that the individual dust particle motion can be precisely analyzed. Thus, various physical procedures can be studied at the individual particle level, such as the phase transition~\cite{Feng:2008}, the diffusion~\cite{Liu:2008}, and the shock propagation~\cite{Kananovich:2020}. Recently, 1D and 2D periodic substrates are introduced to 2D dusty plasmas to modulate their collective structural and dynamical behaviors using simulations~\cite{Li:2018,Huang:2022,Wang:2018,Feng:2021}. In~\cite{Li:2019,Gu:2020}, the 1D-substrate-modulated 2D dusty plasmas are driven by an gradually changing uniform DC force, leading to the pinning and depinnnig dynamics, where the pinned, the plastic flow, and the moving ordered states are observed. However, the directional locking effect has never being investigated in dusty plasmas before. so that it is still not clear whether the directional locking also exists in underdamped dusty plasmas, as studied here. As compared with those overdamped systems~\cite{Korda:2002,MacDonald:2003,Lacasta:2005}, the role of the inertial term in the directional locking effect needs to be further studied.

The rest of this paper is organized as follows. In Sec.\uppercase\expandafter{\romannumeral2}, we describe our simulation method to mimic a substrate-modulated 2D dusty plasma driven by forces at different angles. In Sec.\uppercase\expandafter{\romannumeral3}, we present our observed prominent directional locking effect in a square-substrate-modulated 2D solid dusty plasma, or 2D Yukawa solid, which is driven by the external force in varying directions. We find that the observed prominent locking steps correspond to the flow patterns with the 1D channels along the directions related to the symmetry of the substrate. Around the termini of these locking steps, we investigate the corresponding structure transitions in detail. Finally, we give a brief summary.

\section{Simulation methods}

We follow the tradition~\cite{Morfill:2009,Fortov:2005,Piel:2010,Bonitz:2010} to characterize our studied 2D dusty plasma solids or 2D Yukawa solids using the two dimensionless parameters of the coupling parameter $\Gamma$ and the screening parameter $\kappa$. They are defined~\cite{Ohta:2000,Sanbonmatsu:2001} as $\Gamma = Q^2/(4 \pi \epsilon_0 a k_{B} T)$ and $\kappa = a / \lambda_{D}$, respectively. Here, $Q$ is the charge of one particle, $a = 1/\sqrt{\pi n}$ is the Wigner-Seitz radius~\cite{Kalman:2004} for the areal number density $n$, $T$ is the averaged kinetic temperature from the velocity fluctuation of particles, and $\lambda_D$ is the Debye screening length.   

To investigate the directional locking effect of the substrate-modulated 2D Yukawa solid driven by the external driving force in varying directions, we perform Langevin dynamical simulations~\cite{Li:2018} with $N = 1024$ particles constrained in a rectangular box of $61.1a\times52.9a$ with the periodic boundary conditions. The equation of motion for the $i$th particle is
\begin{equation}\label{Langevin}
{	m \ddot{\bf r}_i = -\nabla \Sigma \phi_{ij} - \nu m\dot{\bf r}_i + \xi_i(t)+{\bf F}_s+{\bf F}_d }.
\end{equation}
The first term on the right-hand side of Eq.~(\ref{Langevin}) is the interparticle Yukawa repulsion~\cite{Konopka:2000} between the $i$th and $j$th particles, where $\phi_{ij} = Q^2 {\rm exp}(-r_{ij} / \lambda_D) / 4 \pi \epsilon_0 r_{ij}$, and $r_{ij}$ is the distance between them. The second term $- \nu m\dot{\bf r}_i$ is the fractional gas damping~\cite{Liu:2003} that particles experience while moving in the rarefied gas or plasma environment. The third term $\xi_i(t)$ is the Langevin random kicks~\cite{Feng:2008,Feng:20082} from the plasma environment, which is realized using the driven-dissipation theorem~\cite{Gunsteren:1982} in our simulations. The last two terms of ${\bf F}_s$ and ${\bf F}_d$ are the forces from the modulate substrate and the external driving source in varying angles, respectively, as we explain in details next.

\begin{figure}[htb]
\centering
\includegraphics{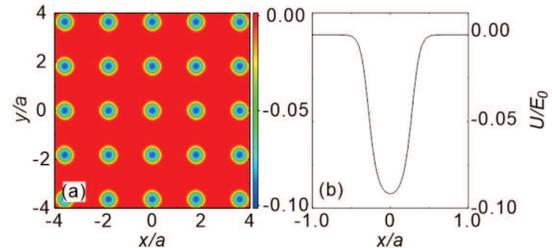}
\caption{\label{substrateU} (a) A contour plot of a portion of the square periodic substrate of Eq.~(\ref{substrate}) for the parameters of $V_{0}=0.1E_{0}$, $w_{x}=1.79a$, $w_{y}=1.82a$, $A=6$, $B=0.8$ in our simulations. (b) A potential profile $U(x,0)$ of one well of the periodic substrate, with the accurate depth of $0.0917E_{0}$.
}
\end{figure}

Our simulated 2D Yukawa solid is modulated by a square periodic substrate, as shown in the Fig.~1(a). As in~\cite{Lacasta:2005}, the function of the substrate is specified as
\begin{equation}\label{substrate}
U(x, y)=\frac{-V_{0}}{1+e^{-g(x, y)}},
\end{equation}
where $g(x,y)$ is the 2D periodic function of $g(x, y)=A\left[\cos \left(2 \pi x / w_{x}\right)+\cos \left(2 \pi y / w_{y}\right)-2 B\right]$~\cite{Lacasta:2005}, which is controlled by five parameters of $V_0$, $w_x$, $w_y$, $A$, and $B$. Clearly, $V_0$ determines the depth of the square substrate in units of $E_0=Q^2/4\pi\epsilon_0 a$, while $w_x$ and $w_y$ are the distances between the centers of two adjacent wells in the $x$ and $y$ directions, respectively, in units of the Wigner-Seitz radius of $a$. Here, $A$ directly controls the steepness of the wells, while the relative width or shape of the well with respect to the spatial periods $w_x$ and $w_y$ is determined by $B$. By taking the derivative of Eq.~(\ref{substrate}), we obtain the force from this substrate as
\begin{equation}\label{substrateforce}
\begin{split}
{\bf F}_{s}=&-\frac{\partial U(x, y)}{ \partial x} \bm{\hat{x}}-\frac{\partial U(x, y)}{ \partial y} \bm{\hat{y}}\\
=&{-\frac{2 \pi A V_{0}}{w_{x}}  \sin \left(\frac{2 \pi x}{w_{x}}\right) \frac{e^{-g(x, y)}}{{\left[1+e^{-g(x, y)}\right]}^{2}}} \bm{\hat{x}}\\
&{-\frac{2 \pi A V_{0}}{w_{y}} \sin \left(\frac{2 \pi y}{w_{y}}\right)  \frac{e^{-g(x, y)}}{{\left[1+e^{-g(x,y)}\right]}^{2}}} \bm{\hat{y}}.
\end{split}
\end{equation}
Our chosen parameters of the square potential function are $V_0=0.1E_0$, $A=6$, $B=0.8$, $w_x=1.79a$, and $w_y=1.82a$. As a result, our simulation box of $61.1a\times52.9a$ contains $N_w = 34 \times 29 = 986$ potential well minima for our specified substrate, leading to the corresponding commensurability ratio of $\rho= N /N_w = 1.039$. In fact, due to the complicated expression of Eq.~(\ref{substrate}), for our specified parameters, the exact depth of each potential well is $0.0917E_0$, not exactly the same as $U_0 = 0.1E_0$, as shown in Fig.~1(b). Note, although we intend to investigate the modulation by the square periodic substrate, our choice of $w_x$ and $w_y$ are slightly different, corresponding to $\approx 1.5\%$ difference in the spatial periods in the $x$ and $y$ directions, to satisfy the periodic boundary conditions for the simulation box of $61.1a\times52.9a$, since we mainly focus on the commensurability ratio $\rho$ of around unity here.

The last term in Eq.~(\ref{Langevin}) is the externally applied driving force ${\bf F}_d=F_x\bm{\hat{x}}+F_y \bm{\hat{y}}$ in units of $F_0=Q^2/4 \pi \epsilon_0 a^2$. For our current investigation, we mainly focus on the dynamical response of the substrate-modulated Yukawa solid to the external driving force in varying directions, so that we keep the external force in the $x$ direction unchanged as $F_x=0.24F_0$, and then only increase the external force in the $y$ direction $F_y$~\cite{Reichhardt:1999} gradually from zero to 0.3$F_0$. As a result, while $F_y$ increases in this range, the corresponding angle of the driving force relative to the $x$ direction $\theta$ varies from ${0}^{\circ}$ to $\approx 51^{\circ}$. Note, for each specified value of the driving force $F_y$ reported here, we always perform one individual simulation run.

Other parameters and the details of our simulations are listed below. The conditions of our simulated 2D Yukawa system are specified as $\Gamma=1000$ and $\kappa=2$, just corresponding to the typical Yukawa solid state~\cite{Hartmann:2005}, so that the temperature effect is reduced. The frictional gas damping rate is chosen as $\nu = 0.027 {\omega}_{pd}$, comparable to the typical 2D dusty plasma experiments~\cite{Feng:2011}, where $\omega_{pd}^{-1}={(Q^2/2 \pi \epsilon_0 m a^3)}^{-1/2}$ is the nominal dusty plasma frequency~\cite{Kalman:2004}. For our simulated Yukawa solid, the time step is specified as $0.0014\omega_{pd}^{-1}$, as well justified in~\cite{Liu:2005}. For each simulation run, after the steady state is achieved, we integrate Eq.~(\ref{Langevin}) for $\ge 5\times{10}^6$ steps for all $N = 1024$ particles to obtain their positions and velocities, which are used for the data analysis reported later. When our simulated system reaches the final steady state, the total collective drift velocity, the total energy, and the kinetic temperature do not change or drift beyond the typical fluctuation level for 1024 particles anymore. Note, we also perform a few test runs with $N = 4096$ particles to make sure that all results presented here are independent from the simulation system size. Other simulation details are the same as in~\cite{Li:2019}.

\section{results and discussions}

\subsection{Locking steps}

To investigate the dynamical response of the square-periodic-substrate-modulated 2D Yukawa solid driven by the uniform force in various angles, we first focus on the collective dynamical measure of the averaged velocity direction, characterized by $tan\psi$. Here, $\psi$ is the collective velocity angle with respect to the $x$ direction. Thus, the value of $tan\psi$ is just $\tan\psi={V_{y}}/{V_{x}}$, where $V_x$ and $V_y$ are the collective velocity projected in the $x$ and $y$ directions, respectively. We calculate $V_x$ and $V_y$ using ${V_{x}=N^{-1}\left\langle\sum_{i=1}^{N}\bm{v_{i}} \cdot \bm{\hat{x}}\right\rangle}$ and ${V_{y}=N^{-1}\left\langle\sum_{i=1}^{N}\bm{v_{i}} \cdot \bm{\hat{y}}\right\rangle}$, respectively. Here, $\bm{v_{i}}$ is the velocity of the particle $i$, and $\left\langle~\right\rangle$ denotes the average for all particles. As the driving force in the $y$ direction $F_y$ increases from zero, we calculate the corresponding $V_x$ and $V_y$ values to determine $tan\psi$ for all conditions, as presented in Fig.~2(a). Note, after each driving force is specified in our simulations, we always monitor the evolution of the total energy and the kinetic temperature of the whole system to confirm that the simulated system arrives at the final steady state, then we start to record the position and velocity data for the calculation of $tan\psi$ reported here.

As the major result of this paper, we discover the prominent directional locking effect from the substrate-modulated 2D Yukawa solid driven by the uniform force in various angles, as presented in Fig.~2(a). As $F_y$ increases from zero, instead of increasing monotonically, $tan\psi$ exhibits several remarkable steps, where the value of $tan\psi$ is completely unchanged. The unchanged value of $tan\psi$ means that the angle of $\psi$ is unchanged while $F_y$ varies, indicating that, for this range of $F_y$, the direction of the collective velocity is ``locked'' in this specific direction, instead of changing with the varying driving force. Interestingly, the locking steps of $tan\psi$ occur when the ratio of $F_y/F_x$ is close to the rational values of $p/q$, where $p$ and $q$ are both integers. Clearly, the most prominent locking step occurs at $p/q = 1/1$, with the $tan\psi$ value of 1. The locking steps at $p/q = 1/2$ and $1/3$ are also both remarkable, corresponding to the $tan\psi$ value of 0.5 and 0.33, respectively. We speculate that these rational values of $p/q = 1/1$, $1/2$, and $1/3$ are related to the locked collective velocities on these steps, probably due to the symmetry direction of the square substrate, as we discuss in detail later.

In Fig.~2(a), for the data beyond the reported steps, we can clearly observe that the value of $tan\psi$ increases linearly with $F_y$ with a uniform slop, as the dashed line shown. The slope of the dashed line is just $1/F_x$, indicating that the increase rate of $tan\psi$ relative to $F_y$ is just $1/F_x$ for our current simulation data. This increase rate of $1/F_x$ means that the value of $tan\psi$ is the same as that of $tan\theta=F_y/F_x$. That is to say, even the studied Yukawa solid is modulated by the square periodic substrate, the observed collective velocity still exactly follows the applied external force direction, for all of the non-locking data in Fig.~2(a).

\begin{figure}[htb]
\centering
\includegraphics{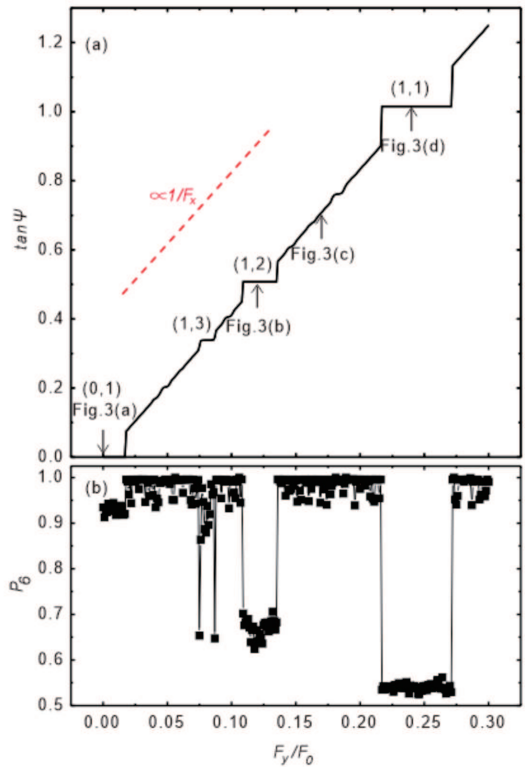}
\caption{\label{locking} Obtained tangent values of the collective velocity angle $\tan\psi$ (a) and the corresponding fraction of sixfold coordinated particles $P_{6}$ (b) versus the transverse driving force $F_{y}$ for the square-substrate-modulated 2D Yukawa solid, with the commensurability ratio of $\rho=1.039$. Here the driving force $F_{x}$ is specified as an unchanged value of $F_{x}=0.24F_{0}$. Locking steps occur when $F_{y}/F_{x}$ is near rational values of $F_{y}/F_{x}=p/q$, where $p$ and $q$ are both integers. The locking steps $(p, q)$ of (0, 1), (1, 3), (1, 2), (1, 1) are manifested in (a), where (1, 1) is the most prominent. Corresponding to these locking steps, substantial diminishments in the structural measure of $P_{6}$ are clearly exhibited in (b), especially for the significant drops in $P_{6}$ on the (1, 2) and (1, 1) locking steps. At the termini of the (1, 3) locking step, there are two abrupt dips in $P_{6}$.
}
\end{figure}

From Fig.~2(a), as $F_y$ increases gradually from zero, the value of $tan\psi$ is always zero when $F_y \lesssim 0.018F_0$, which can be termed as the (0, 1) locking step, as named in~\cite{Reichhardt:1999,Reichhardt:2012}. On the (0, 1) locking step, the $tan\psi$ value of zero indicates that the Yukawa solid is completely pinned in the $y$ direction, although all particles move in the $x$ direction, i.e., ``locked'' in the $x$ direction. When $F_y\ge 0.018F_0$, from Fig.~2(a), the $tan\psi$ value abruptly increases from zero to the linear straight line of $tan\psi=F_y/F_x$. As in~\cite{Reichhardt:1999}, the value of $F_y=0.018F_0$ here can be termed as the critical transverse depinning force $F_y^c$ for our studied system.

In fact, besides the (1, 1), (1, 2), (1, 3), and (0, 1) locking steps reported above, we can still observe some small ripples in the linear increase portion of $tan\psi$ in Fig.~2(a). For example, these ripples occur around $F_y/F_0 = 0.048$, 0.097, and 0.182. Probably these ripples are also related to other rational values of $p/q$. However, due to the finite step of $F_y$ in our simulations and the finite accuracy of the square periodic substrate, here we do not explore these small ripples further, as discussed later.

To further analyze the mechanism of the observed directional locking, besides the dynamical diagnostic of $tan\psi$, we also investigate the structure of our studied 2D Yukawa solid during the depinning procedure, as shown in Fig.~2(b). To characterize the structure, we calculate the fraction of sixfold coordinated particles $P_6$~\cite{Reichhardt:2005} using $\textstyle{P_{6}=N^{-1}\left\langle\sum_{i=1}^{N} \delta\left(6-z_{i}\right)\right\rangle}$, where $z_i$ is the coordination number of the $i$th particle obtained in the Voronoi construction~\cite{Li:2019}. Note, for a perfect 2D triangular lattice, $P_6 = 1$, while for a more disordered state, the value of $P_6$ is smaller. From the calculated results of $P_6$ in Fig.~2(b), clearly, the value of $P_6$ keeps at a relatively high value, mostly larger than 0.95, during the non-locking phase, i.e., the linear increasing portion of $tan\psi$. This feature indicates that our studied 2D Yukawa solid exhibits the highly ordered triangular arrangement outside these locking steps.

Our observed directional locking effect from $tan\psi$ is also distinctively exhibited from the structural measurement of $P_6$ in Fig.~2(b). For the most prominent (1, 1) locking step, the $P_6$ results exhibit a significant drop to the steady low value of $\approx 0.52$. The $P_6$ value also drops to $\approx 0.67$ for the remarkable (1, 2) locking step. However, for the (1, 3) locking step, the $P_6$ results exhibit two abrupt dips to $\approx 0.65$ then immediately restoring back to the high values of around 0.9. That is to say, unlike the (1, 1) and (1, 2) locking steps where the $P_6$ value always stays at the lower values of $\textless~0.7$, the $P_6$ value only momentarily drops below 0.7 at the two termini of the (1, 3) locking step, while at the central portion of the (1, 3) locking step, the $P_6$ value is still high $\approx 0.9$. For the (0, 1) locking step, although the $P_6$ value is still high $\approx 0.93$, a distinctive step slightly lower than the non-locking phase can be clearly identified.

\subsection{Flow patterns}

To illustrate the detailed dynamics of the directional locking effect, in Fig.~3 we plot the typical particle trajectories in various conditions as marked in Fig.~2(a). In Fig.~3, we use filled circles to indicate the particle positions in one typical snap shot, and then use open circles to label the locations of potential wells of the square periodic substrate. The straight lines shown in Fig.~3 are just the typical particle trajectories during the time interval of $2.8{\omega}_{pd}^{-1}$. The four conditions in each panel of Fig.~3 correspond to the (0, 1) locking step, the (1, 2) locking step, the non-locking phase of $F_{y} = 0.17F_{0}$, and the (1, 1) locking step, respectively, as marked in Fig.~2(a).

The particle trajectories exhibit the distinctive feature in the locking step conditions in Fig.~3. For the most prominent (1, 1) locking step, all particles move regularly along the 1D channels at the same direction of $\psi = {45}^\circ$, as shown in Fig.~3(d). As shown in Fig.~2(a), even the value of $F_y / F_0$ varies in the range of $[0.217, 0.271]$ , $tan\psi$ always stays at the constant value of unity, indicating that the $\psi={45}^\circ$ 1D channels are the only choice of the particle trajectories, i.e., the direction with the angle of $\psi={45}^\circ$ is ``locked''. Similarly, for the (1, 2) locking step, all particles also move regularly along the 1D channels at the same direction of $\psi=tan^{-1}{(1/2)}$, as shown in Fig.~3(b), so that the particle motion distance in the $x$ direction is always twice of that in the $y$ direction. In fact, for the (1, 3) locking step, we also verify that all particles move along 1D channels at the direction of $\psi=tan^{-1}{(1/3)}$. In short, we find that all of the observed locked directions of the particle motion are along the symmetric directions of the square periodic substrate.

As shown in Fig.~3(a), for the (0, 1) locking step, the corresponding particle trajectories are always along the $x$ direction, which are just the 1D channels for the particle motion. That is to say, all particles are still completely pinned in the $y$ direction. When the applied force $F_y$ exceeds the critical transverse depinning force $F_y^c$~\cite{Reichhardt:1999}, particles are able to move in the $y$ direction, so that the depinning procedure in the $y$ direction occurs. Note, as in~\cite{Reichhardt:1999,Gopinathan:2004}, the stable directions of the particle motion $\psi$ corresponding to these locking steps shown in Figs.~3(a, b, d) are often termed as the commensurate angles.

For the non-locking phase conditions, such as $F_{y} = 0.17F_{0}$ as shown in Fig.~3(c), the particle trajectories are completely different from those on the locking steps. As shown in Fig.~3(c), the regular 1D channels completely disappear, however, the particle trajectories are nearly uniformly distributed, exhibiting the quasiperiodic feature, similar to those observed in~\cite{Reichhardt:1999}. The directions of the particle motion $\psi$ corresponding to these non-step conditions may be termed as the incommensurate angles. Furthermore, from Fig.~3(c), it is clear that these particles are self-organized in a highly ordered triangular lattice, well agreeing with the corresponding high value of $P_6$ in Fig.~2(b).

\begin{figure}[htb]
\centering
\includegraphics{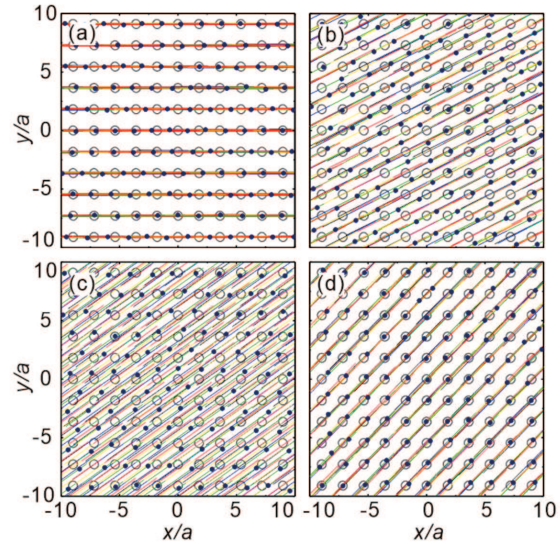}
\caption{\label{trajectories} Obtained particle trajectories of the 2D Yukawa solid for the time duration of $2.8{\omega}_{pd}^{-1}$ at the conditions corresponding to the (0, 1) locking step (a), the (1, 2) locking step (b), the (1, 1) locking step (d), and the non-locking condition of $F_{y} = 0.17F_{0}$ (c), as marked in Fig.~2(a). The large open circles present the locations of potential wells of the substrate, while the small filled circles indicate the particle positions at one typical snap shot. From these trajectories, at the locking steps, particles mostly move in 1D channels aligned with potential wells as in (a, b, d). While under the non-locking condition, particle trajectories exhibit the quasiperiodic feature as in (c). On the (0, 1) locking step in (a), all particles are completely pinned in the $y$ direction as they move in the $x$ direction, i.e., locked in the $x$ direction. On the (1, 1) locking step in (d), particles always move along the diagonal direction with the same moving distance in the $x$ and $y$ directions, while on the (1, 2) locking step in (b), the moving distance in the $x$ direction is twice of that in the $y$ direction. Note, different colors of the particle trajectories correspond to different times.
}
\end{figure}

\subsection{Structures}

To further characterize the different structures corresponding to these locking steps and the non-locking phase, in Fig.~4 we also calculate the 2D distribution functions $G_{xy}$ of the studied system in the same four conditions as in the four panels of Fig.~3. The 2D distribution function $G_{xy}$~\cite{Loudiyi:1992} provides the probability to find a particle at one position relative to the central reference particle, which is often used for anisotropic systems. While calculating $G_{xy}$, for each applied driving force, we also use the simulated particle position data in the final steady state, similarly to the calculation of $tan\psi$ above. For each panel of Fig.~4, besides $G_{xy}$, we also provide a typical snap shot of the corresponding particle positions on the lower left corner to illustrate the arrangement of particles.

From the obtained four $G_{xy}$ results in Fig.~4, clearly, the structure of Fig.~4(c) is the most ordered triangular lattice with the hexagonal symmetry. Besides Fig.~4(c), we also confirm that the particle arrangement corresponding to these non-step phase conditions, or the incommensurate angles, is always highly ordered. The corresponding snap shot further verifies the highly ordered triangular lattice in these conditions. In the depinning procedure of the 1D-periodic-substrate-modulated 2D Yukawa systems~\cite{Li:2019,Gu:2020}, when the external driving force is large enough, the system is in the moving ordered state, i.e., all particles move in a highly ordered triangular lattice as one body. From our understanding, in the non-locking phase conditions as Fig.~4(c), the studied 2D Yukawa solid is also in the moving ordered state, similar to those in~\cite{Li:2019,Gu:2020}.

\begin{figure}[htb]
\centering
\includegraphics{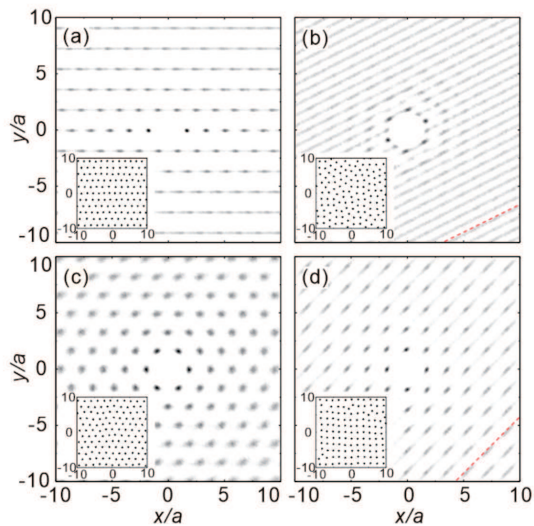}
\caption{\label{Gxy} Obtained 2D distribution functions $G_{xy}$ for the substrate-modulated 2D Yukawa solid at the same conditions as in Fig.~3. For all these four conditions, the non-locking condition (c) corresponds to the most ordered triangular lattice, while the (1, 2) locking step in (b) corresponds to the slightly disordered structure. The (0, 1) step in (a) corresponds to the triangular lattice with the main axis along the $x$ direction. The (1, 1) locking step in (d) almost corresponds to the square lattice. For each panel, the inset on the lower left corner is the typical snapshot of particle positions.
}
\end{figure}

The calculated 2D distribution functions $G_{xy}$ on these locking steps exhibit the completely different features in Fig.~4. For the most prominent (1, 1) locking step, from both $G_{xy}$ and the corresponding snap shot in Fig.~4(d), the particles are arranged to a nearly square lattice. We speculate that this square lattice arrangement is probably caused by the combination of two facts, which are the nearly same magnitude of the driving forces in the two directions and the nearly same period of the potential well arrays in the two directions. From our understanding, all particles almost experience the same magnitude of force in the $x$ and $y$ directions, not only modulated by the square substrate, but also driven by the external force with the direction of $\theta \approx {45}^\circ$. The forces from the external driving force and the substrate are both substantially higher than the interparticle force, as a result, the nearly same magnitude of the external forces in the $x$ and $y$ directions reasonably result in the square lattice in Fig.~4(d). Further away from the center of $G_{xy}$, the points related to the distributed particle positions seem to be smashed along the direction of $\psi={45}^\circ$, as the dashed line shown in the lower right corner of Fig.~4(d). We speculate that this smashed feature in $G_{xy}$ is caused by the particle motion along the 1D channels, as shown in Fig.~3(d).

For the (0, 1) locking step, the calculated $G_{xy}$ and the corresponding snap shot in Fig.~4(a) further verify that all particles are pinned in the $y$ directions. Particles only move in the $x$ direction, so that they are well aligned in the $y$ direction. From the calculated $G_{xy}$, the particles are mainly ordered in the $x$ direction, while in the $y$ direction, the particles in different lines are not very correlated. From the calculated $G_{xy}$ in Fig.~4(a), the points far away from the center also seem to be smashed along the $x$ direction. Clearly, this smashed feature along the $x$ direction should be caused by the particle motion only in the $x$ direction.

For the (1, 2) locking step, the calculated $G_{xy}$ and the corresponding snap shot in Fig.~4(b) both exhibit much more disordered feature than Figs.~4(a, c, d). From $G_{xy}$ in Fig.~4(b), except for the nearest neighbors of the center, other points related to the particle positions are smashed along the direction of $\psi=tan^{-1}{(1/2)}$, as the dashed line shown in the lower right corner of Fig.~4(b). As in Figs.~4(a, d), this smashed feature in $G_{xy}$ should be caused by the particle motion along the 1D channels, as shown in Fig.~3(b). For the current condition, $F_y$ is only about one half of $F_x$, so that the driving force in the $x$ and $y$ directions are not commensurate to induce the square lattice as in Fig.~4(d). In fact, we may regard the structure of Fig.~4(b) as the combination of Figs.~4(a) and 4(d).

\begin{figure}[htb]
\centering
\includegraphics{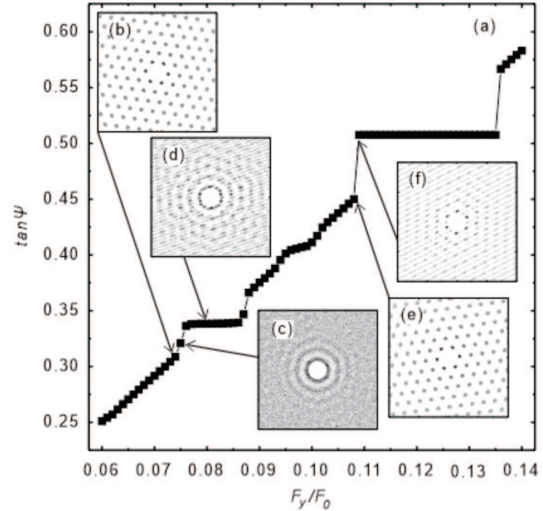}
\caption{\label{Details} Magnified view of $\tan\psi$ (a) versus the transverse driving force $F_{y}$ as in Fig.~2(a), and the corresponding 2D distribution functions $G_{xy}$ (b, c, d, e, f) around the termini of the two locking steps. The $G_{xy}$ results in (b) and (e) both correspond to the highly ordered triangular structure, while the $G_{xy}$ results in (d) and (f) correspond to the structure with the smashed feature along the direction of $\psi=tan^{-1}{(1/3)}$ and $\psi=tan^{-1}{(1/2)}$, respectively. The $G_{xy}$ results in (c) indicate the highly disordered structure, i.e., the disordered plastic flow state. At the left terminus of the (1, 3) locking step, $\tan\psi$ increases from the non-locking phase (b) to the disordered plastic flow state (c) and then to the (1, 3) locking phase (d) finally. Similarly, at the right terminus of the (1, 3) locking step, $\tan\psi$ increases also from the locking phase to the disordered plastic flow state and then to the non-locking phase. However, for the (1, 2) locking step, the transition between the non-locking phase (e) and locking phase (f) is discontinuous, without the disordered plastic flow state between them.
}
\end{figure}

To further investigate the dynamical and structural transitions around the termini of the locking steps, we magnify one portion of the $tan\psi-F_y$ results in Fig.~5(a). In Fig.~5, we also plot five 2D distribution functions, corresponding to five different conditions around the termini of the (1, 3) and (1, 2) locking steps. From Fig.~5(a), clearly, the $tan\psi$ exhibits abrupt jumps between the linear increase non-locking phase and the two termini of the (1, 2) locking step, indicating the discontinuous transition between the non-locking and the locking phases there. For the non-locking phase, the obtained 2D distribution function $G_{xy}$ in Fig.~5(e) clearly indicate a highly ordered triangular lattice, well consistent with the moving ordered state driven at unlocked angles, as in Fig.~4(c). When $tan\psi$ reaches the left terminus of the (1, 2) locking step, the calculated 2D distribution function $G_{xy}$ in Fig.~5(f) clearly exhibits the smashed feature along the direction of $\psi=tan^{-1}{(1/2)}$, exactly the same as Fig.~4(b). That is to say, for all conditions on the locking step, the structure does not change any more, just like the locked moving direction of $\psi=tan^{-1}{(1/2)}$. At the right terminus of the (1, 2) locking step, the 2D distribution function $G_{xy}$ also abruptly changes from the smashed feature along $\psi=tan^{-1}{(1/2)}$ to the highly ordered triangular lattice. In addition, we confirm that the phase transition between the non-locking phase and the (1, 1) locking step, or the (0, 1) locking step, is also discontinuous, very similar to the transition related to the (1, 2) locking step reported above.

From Fig.~5, we observe a different phase transition between the non-locking phase and the (1, 3) locking step, which is associated with the disordered plastic flow state. In Fig.~5(a), between the non-locking phase and the (1, 3) locking step, we find that there are a few data points corresponding to the angles which are neither the direction of the driving force nor the locked angle of $\psi=tan^{-1}{(1/3)}$, such as the one marked with the arrow of Fig.~5(c). To find the corresponding structure transition, we calculate 2D distribution functions $G_{xy}$ on the conditions of the three data points around the left terminus of the (1, 3) step, as shown in Figs.~5(b-d). In Fig.~5(b), the hexagonal symmetric 2D distribution functions $G_{xy}$ indicates the highly ordered triangular lattice, just corresponding to the typical moving ordered state in the non-locking phase as Fig.~4(c). At the left terminus of the (1, 3) locking step in Fig.~5(d), the 2D distribution functions $G_{xy}$ shows some disordered feature with the smashed feature related to the 1D moving channels along the direction of $\psi=tan^{-1}{(1/3)}$, similar to Fig.~5(f). However, for the conditions between Figs.~5(b) and 5(d), the calculated $G_{xy}$ clearly exhibits the ring-shaped 2D distribution in Fig.~5(c), indicating the arrangement of particles is much disordered as a liquid. That is to say, for the transition from the non-locking phase to the (1, 3) locking step, our studied 2D Yukawa solid changes from the moving ordered state to the disordered plastic flow state first, then to the locking phase when $tan\psi$ arrives at the left terminus of the (1, 3) locking step. Around right terminus of the (1, 3) locking step, this studied system changes from the locking phase to the disordered plastic flow state first, and then to the moving ordered state of the non-locking phase as the $\tan\psi$ value increases. In fact, this transition feature is also verified from the momentarily drops of the structural measure of $P_6$ at the two termini of the (1, 3) locking step in Fig.~2(b), which is completely different from other locking steps.

\subsection{Discussions}

Under the conditions of the observed locking steps, it is clear that all particles move along 1D channels aligned with the directions corresponding to the symmetry of the substrate, as shown in Fig.~3. Here, we provide our interpretation of the observed locking steps. For the most prominent (1, 1) locking step, the particles all move through potential wells along the $\psi={45}^\circ$ diagonal direction of the square substrate, as shown in Fig.~3(d). When $F_y = F_x$, as the particle drifts from one potential well, it falls to the next potential well in the diagonal direction soon. Even if $F_y$ is mismatched with $F_x$ not too much, after the particle leaves the initial potential well, it is still captured by the next potential well in the diagonal direction. Thus, even for the mismatched $F_y$ and $F_x$, the moving direction of particles is still locked in the $\psi={45}^\circ$ direction.

Following on this interpretation, the $(p, q)$ locking steps correspond to the symmetric drift directions, where particles move $q w_x$ in the $x$ direction while it moves simultaneously $p w_y$ in the $y$ direction. When the values of $p$ and $q$ are smaller, such as either 1 or 0, as particles move from one well to the next one, they do not pass other or many potential wells, so that this direction is easily ``locked'' even if the forces in two directions are somewhat mismatched, i.e., the locking effect is prominent, like the (1, 1) and (0, 1) locking steps. However, when the values of $p$ and $q$ are larger, as particles move from one well to the next one, they pass much more other potential wells, so that particles may be easily captured by these wells close to their trajectories if the forces in two directions are slightly mismatched, i.e., the locking effect is not very prominent, like the (1, 3) locking step, well consistent with the results in Fig.~2.

Here, we would like to clarify that all of the conclusions presented above are based on our current Langevin simulations. Computer simulations always contain some limitations, for example, the finite number of discrete data points. Our Langevin simulations presented here contain the finite resolution due to at least three points. First, while increasing the transverse force $F_y$, we choose the finite increasing step of $0.001F_0$. Our discontinuous transition between the non-locking phase and the (1, 1) or (1, 2) locking step is drawn based on this $0.001F_0$ increasing step of $F_y$. Second, for our current simulation box, due to the different scales in the $x$ and $y$ directions, to satisfy the periodic boundary conditions, the distances between the centers of two adjacent wells of the square substrate in the two directions are set as $1.79a$ and $1.82a$, with $\approx 1.5\%$ difference, i.e., the square substrate is not perfectly square. Third, the coupling parameter of our studied 2D Yukawa solid is $\Gamma=1000$, so that the corresponding thermal energy may introduce the stochastic feature in the dynamical diagnostics. These latter two points would both blur the locking step feature in Fig.~2. Due to the precision resolution problem caused by these points, it is quite difficult to observe the locking steps with larger values of $p$ and $q$ using our current simulation method.

\section{Summary}

In summary, we investigate the directional locking effect in a square-periodic-substrate-modulated 2D Yukawa solid driven by the external force with varying directions. When the direction of the driving force is close to the symmetry directions of the square substrate, we find four distinctive directional locking steps, characterized by the diagnostic of $tan\psi$, where the direction of the whole collective motion of all particles $\psi$ is locked. Under the conditions of each locking step, all individual particle move regularly along the 1D channels at the same direction related to this locking step. The non-locking phase always corresponds to the most ordered triangular lattice with the hexagonal symmetry, while the structure of our system in the locking phase depends on the corresponding locking step. For the most prominent (1, 1) locking step, the system exhibits the distinctive square lattice arrangement. In addition, we also find the transition between locking and non-locking phases is either discontinuous for some locking steps, such as the (1, 1) and (1, 2) locking steps, or continuous for others like the (1, 3) locking step. Around the termini of the (1, 3) locking step, while the system changes between the moving ordered non-locking phase and the locked phase, the disordered plastic flow state always occurs, which is completely different from the (1, 1) and (1, 2) locking steps. We also provide our interpretation of the observed directional locking locking.

The previous directional locking studies are mainly focused on overdamped systems, while our current investigation shows that underdamped 2D Yukawa systems also exhibit the directional locking effect. This study probably is able to open up new application possibilities for underdamped systems moving over periodic substrates, such as sorting different species of dust particles in dusty plasmas.

\section*{ACKNOWLEDGMENTS}
The work was supported by the National Natural Science Foundation of China under Grant Nos. 12175159 and 11875199, the 1000 Youth Talents Plan, startup funds from Soochow University, and the Priority Academic Program Development of Jiangsu Higher Education Institutions, and the U. S. Department of Energy through the Los Alamos National Laboratory. Los Alamos National Laboratory is operated by Triad National Security, LLC, for the National Nuclear Security Administration of the U. S. Department of Energy (Contract No. 892333218NCA000001).



\end{document}